\documentclass[prb,twocolumn,showpacs,amsmath,amssymb,superscriptaddress]{revtex4}
\makeatletter
\usepackage{graphicx}
\usepackage{dcolumn}
\usepackage{bm}

 \def\pto{PbTiO$_3$}

\begin{document}

\title{First-principles study of high-field piezoelectricity
  in tetragonal \pto}

\author{Anindya Roy}
\email{anindya@physics.rutgers.edu} \affiliation{ Department of
  Physics \& Astronomy, Rutgers University, Piscataway, NJ 08854-8019,
  USA }

\author{Massimiliano Stengel}
\affiliation{CECAM - Centre Europ\'een de Calcul Atomique et Mol\'eculaire,
Ecole Polytechnique F\'ed\'erale de Lausanne, 1015 Lausanne, Switzerland}

\author{David Vanderbilt}
\affiliation{ Department of Physics \& Astronomy, Rutgers University,
  Piscataway, NJ 08854-8019, USA }

\date{\today}

\begin{abstract}
We calculate from first principles the nonlinear piezoelectric
response of ferroelectric PbTiO$_3$ for the case
of a polarization-enhancing electric field applied along the
tetragonal axis.  We focus mainly on the case of fixed in-plane
lattice constants, corresponding to epitaxially constrained
thin films.  We find that the dependence of the $c/a$ ratio on
electric field is almost linear in the range up to 500 MV/m, with
little saturation.  This result contrasts with expectations
from Landau-Devonshire approaches based on experimental results
obtained at lower fields, but is in qualitative agreement with a
recent experiment in which higher fields were attained using
pulsed-field methods.  We also study cases in which the in-plane
epitaxy constraint is removed, or an artificial negative pressure is
applied, or both.  These calculations demonstrate that PbTiO$_3$ can
show a strikingly non-linear piezoelectric response under modified
elastic boundary conditions.

\end{abstract}

\pacs{77.65.Ly, 77.84.Dy, 77.65.Bn, 63.20.dk}

\maketitle

\section{Introduction}
\label{sec:intro}

Piezoelectricity was discovered in the late nineteenth century and
has been successfully described at the phenomenological macroscopic level
since the early twentieth century.
Piezoelectric materials find a variety of technological uses including sonar
detection, electromechanical actuators and resonators, and high-precision
applications such as in microbalances and scanning-probe microscopes.
Most of these conventional applications depend on the linearity of the
strain response to the applied electric field or vice versa.
Application of large strains or strong electric fields may,
however, generate a nonlinear response.  Such nonlinearities
are less well characterized and understood, despite the fact
that they may have important practical implications for degrading
or improving device performance.

The intrinsic nonlinear piezoelectric properties of a material may
be difficult to access experimentally for two reasons.  First,
application of large stresses or strains may cause cracking or
other forms of mechanical failure, while application of strong
electric fields may cause dielectric breakdown, leading in
either case to a limited ability to access the nonlinear regime.
Second, most strong (and strongly nonlinear) piezoelectrics are
ferroelectrics, whose spontaneous polarization is strongly modified
by applied strains or fields.  Real samples of ferroelectric
materials usually break up into domains having different directions
of the spontaneous polarization, and the observed piezoelectric
response may be dominated by extrinsic processes associated with
changes in domain populations under applied strain or field.
Studies of single-domain samples are difficult; even if such a
sample is obtained, the domain may tend to reorient under applied
strain or field.  In the case of a strong applied electric
field, one expects that the sample will eventually be poled
into a single-domain state, so that a subsequent increase of the
field strength will access the intrinsic piezoelectric response.
However, one must still overcome the problems associated with
dielectric breakdown mentioned earlier.

In the last two decades or so, predictive quantum-mechanical
descriptions based on first-principles electronic-structure
calculations have been successfully developed and
applied to piezoelectric and ferroelectric materials (see
Ref.~[\onlinecite{rabe-chapter07}] for a recent review).
These methods allow one to isolate the intrinsic contributions
to the nonlinear piezoelectric behavior of a material by calculating
the response explicitly.  We apply such methods here to the study of
tetragonal PbTiO$_3$ (PT) thin films under a strong electric field
applied parallel to the polarization direction.

Our work is motivated in part by recent experimental and modeling
studies of PT and PbZr$_x$Ti$_{1-x}$O$_3$ (PZT) thin
films.\cite{nagarajan-apl02,chen-jap03,grigoriev-prl08}
Based on experimental measurements\cite{nagarajan-apl02} on PT and PZT
films in fields up to $\sim$50\,MV/m,
Landau-Devonshire models were developed to systematize the data
and extrapolate to higher fields.\cite{chen-jap03} For pure PT,
these papers show $d_{33}$ dropping by about 16\% and 10\% as the
electric field is increased to 50\,MV/m for the cases of free-stress
and epitaxially-constrained samples respectively.  (For the latter,
the in-plane lattice constant was constrained to that of the SrTiO$_3$
substrate.)
These results are roughly in agreement
with previous results of a first-principles calculation\cite{sai-prb02}
for the free-stress case, although the methods used there involved
approximations that are removed in the present work.

More recently, Grigoriev {\it et al.}\cite{grigoriev-prl08} measured
the piezoelectric response of epitaxially constrained PZT 20/80 (i.e.,
$x$=0.2) films using a novel approach in which dielectric breakdown was
avoided by the use of ultrashort electric-field pulses in a thin-film
geometry.  This allowed access to much larger electric fields than
previously possible, up to about 500\,MV/m compared with $\sim$50\,MV/m
studied in the earlier experiments.  The strain response
showed almost no saturation up to the highest fields reached in the
experiment, around 500\,MV/m, contrasting with the conclusions of
Refs.~[\onlinecite{nagarajan-apl02,chen-jap03}].  These results may
suggest that the previous work, limited as it was to smaller fields and
longer time scales, may have been more sensitive to extrinsic effects
such as incomplete poling of the domains.

Clearly, information from predictive first-principles calculations would
be very useful here.  A preliminary application of such methods to the
case of stress-free PT under strong applied electric fields
\cite{stengel-np09} showed not only a lack of saturation,
but even an enhancement of the piezoelectric response, with
$d_{33}$ increasing up to fields of about 550\,MV/m.
This unexpected~\cite{nagarajan-apl02,chen-jap03} enhancement was
ascribed to a field-induced structural transition in which the PT crystal
adopts a supertetragonal state at high fields, with an extreme axial
$c/a$ ratio of 1.2-1.3.
Such a supertetragonal state was first identified theoretically
by Tinte {\it et al.},\cite{tinte-prb03} who predicted its
occurrence in PT under conditions of fictitious negative pressure.
Interestingly, a similarly strong tetragonal distortion was recently
reported for other materials, e.g.\ BiCoO$_3$ in its bulk ground
state~\cite{Belik:2006} and BiFeO$_3$ films under a compressive epitaxial
strain.\cite{Bea_et_al:2009}
These results suggest that the peculiar electromechanical response of
PT reported in Ref.~\onlinecite{stengel-np09} might be a rather general
property of many perovskite materials.
In particular, it is not unreasonable to think that, by understanding the
interplay of composition, epitaxial strain, pressure and applied external
fields, one might be able to \emph{control} the crossover between the
competing ``normal'' and ``supertetragonal'' phases in a given Pb- or
Bi-based compound.
This might produce exciting and unusual effects, such as giant piezoelectric
responses and radical magnetoelectric couplings (for example, a magnetic
moment collapse and insulator-to-semimetal transition under pressure was
recently reported for BiCoO$_3$~\cite{Ming}), with a wide range of
implications for technology and fundamental science.

To help shed light on the above issues, in this work we extend the analysis of
Ref.~\onlinecite{stengel-np09} to study the impact of the elastic boundary
conditions on the electromechanical response of PT.
In particular, we address here the case of epitaxially constrained or
stress-free films, with or without external pressure, and
in a wide range of applied electric fields.
For the case of an epitaxially constrained film, our
first-principles calculations indicate that $d_{33}$ of PT monotonically
decreases
with field, but only slowly: by about 2-3\% up to 50\,MV/m, 11-14\% at
500\,MV/m, and 40-50\% at 1500\,MV/m.
Our results thus suggest that, based on intrinsic material properties,
the strain enhancement can continue to quite large fields, with
only a very slow saturation of $d_{33}$, at variance with previous
expectations\cite{nagarajan-apl02,chen-jap03} and strengthening
the findings in Ref.~\onlinecite{grigoriev-prl08}.  In the absence
of an epitaxial constraint, we confirm the previous results of
Ref.~[\onlinecite{stengel-np09}], finding a maximum of $d_{33}$ around
550\,MV/m
and then a gentle drop-off to about a third of its highest value
at 1400\,MV/m. We further investigate the field-induced strain response
of the system under the application of a fictitious negative pressure.

In comparing our work with experimental studies, it is useful to keep in
mind the context and implicit assumptions of our calculations.  Following
the standard framework of density-functional theory, we treat pure
defect-free PT in perfect tetragonal symmetry at zero temperature, with
and without in-plane epitaxial constraint.  We preserve the primitive
periodicity, so that domain formation is excluded.  We study the response
of the system on a time scale on which lattice and strain response can
occur, but domain dynamics cannot occur.  We can go to much larger
fields than can be accessed even by the pulsed-field methods,
\cite{grigoriev-prl08} limited only by the intrinsic breakdown fields
associated with our method for treating the electric field.
\cite{iniguez-prl02,umari-prl02}  We thus study a regime
of intrinsic behavior that, we believe, is closer to
that of Ref.~[\onlinecite{grigoriev-prl08}] than that of
Refs.~[\onlinecite{nagarajan-apl02,chen-jap03}].

The paper
is organized as follows.  We briefly describe the
computational details in Sec.~\ref{sec:methods}. We then present the
results of our calculations in Sec.~\ref{sec:results} and compare them
with experiment.  We also describe further investigations of the
system as we go beyond the experimental conditions, and discuss the
implications of the work.  We then summarize in
Sec.~\ref{sec:summary}.

\section{Computational methods}
\label{sec:methods}

The calculations are performed using density-functional theory
with two {\it ab-initio} computer code packages, ABINIT \cite{ABINIT}
and LAUTREC.\cite{LAUTREC}
We use the Ceperley-Alder\cite{ceperley-prb78,ceperley-prl80}
exchange-correlation, implemented in the
Perdew-Zunger\cite{perdew-prb81} and
Perdew-Wang\cite{perdew-prb92} parameterizations for ABINIT and
LAUTREC respectively.
For ABINIT we use norm-conserving pseudopotentials generated using
the method of Ramer and Rappe\cite{ramer-prb99} as implemented in
the OPIUM package,\cite{OPIUM} while in
LAUTREC we use projector augmented-wave (PAW) potentials.
\cite{blochl-prb94}
In both cases, the semicore $3s$ and $3p$ orbitals of Ti, and
the $5d$ orbitals of Pb are treated as valence electrons.
Plane-wave cutoffs of 50 and 30 Hartree are chosen for
ABINIT and LAUTREC respectively (the PAW potentials
being softer than the norm-conserving ones).  The Brillouin zone is
sampled by a $4\times4\times4$ Monkhorst-Pack \cite{monkhorst-prb76}
$k$-point mesh for ABINIT and a $6\times6\times6$ mesh
for LAUTREC.
A stress threshold of 2$\times$10$^{-2}$~GPa is used for cell
relaxation,\cite{explan-pulay} and forces on ions are converged
below 2.5$\times$10$^{-3}$~eV/\AA.

In ABINIT the electric polarization is calculated using the Berry-phase
approach \cite{king-smith-prb93} and is coupled to a fixed electric field
$E$. \cite{souza-prl02,stengel-prb07}
In LAUTREC the electric polarization is computed using the centers
of the ``hermaphrodite'' Wannier orbitals \cite{stengel-prb06,stengel-prb07}
and the electric displacement field $D$ is used as the independent electrical
variable.\cite{stengel-prb07,stengel-np09}
In either case, the appropriate $E$ or $D$ field is applied, and the
internal coordinates and unconstrained lattice constants are allowed to
relax within the constraints of the enforced {\it P4mm} tetragonal
symmetry.
In both cases, the results are presented as a function of
$E$; for the case of LAUTREC, this is determined
at each $D$ from the computed polarization using
$E = D - 4 \pi P$.

As we shall see, the results obtained using ABINIT and LAUTREC are
generally consistent, but with some quantitative differences
between them.  Tests indicated that the codes produce almost
identical results when using the same potentials, so we can ascribe
these differences almost entirely to the different potentials used
(norm-conserving for ABINIT vs.\ PAW for LAUTREC).

\section{Results}
\label{sec:results}

\subsection{\label{sec:cal-expt}Piezoelectric response with epitaxy constraint}

\begin{figure}
\includegraphics[width=3.4in]{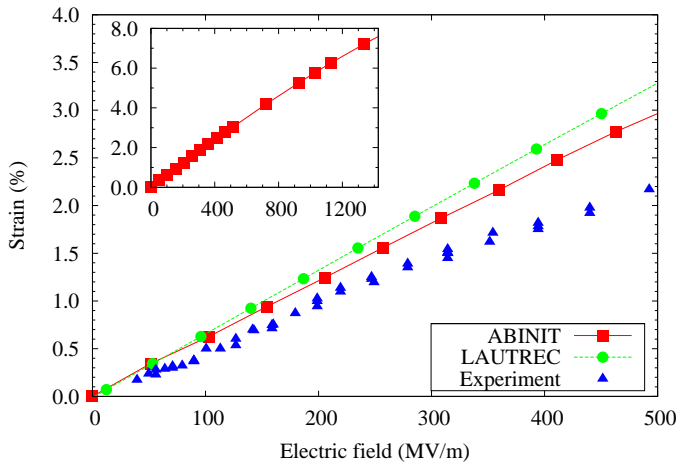}
\caption{\label{fig:comp-comp}
  Out-of-plane strain response to applied electric field,
  relative to zero-field ferroelectric state, under in-plane
  epitaxial constraint.  Experiment: data of Grigoriev {\it
  et al.} \cite{grigoriev-prl08} for several PZT 20/80 capacitors
  of varying thickness.  Theoretical points with fitted lines:
  computed results for pure PbTiO$_3$ using two computer codes.
  Inset shows the ABINIT results over a more extended range
  of fields.}
\end{figure}

Both the ABINIT and LAUTREC computer codes were used to compute the
response of the PT system to electric fields up to 1400\,MV/m
applied parallel to the polarization (along the tetragonal axis).
The epitaxy constraint was enforced by fixing the in-plane lattice
constant $a$ to remain at a value 1.14\% smaller than the computed
equilibrium $a_0$ of bulk tetragonal PT; this factor was chosen
to facilitate comparison with the experiments of Grigoriev {\it
et al.}\cite{grigoriev-prl08}
in which the PZT 20/80 film was compressively strained by 1.14\%
relative to its bulk equilibrium in-plane lattice constant
by its epitaxial coherence with the SrTiO$_3$ (ST) substrate.
Having computed the equilibrium $a_0$ to be 3.892 and 3.854\,\AA\ using
ABINIT and LAUTREC respectively, we therefore
set the constrained $a$ to be 3.848 and 3.811\,\AA\ for
the two codes respectively.  (We have traced the shift of lattice
constant between codes to the choice of Ti pseudopotential, but
trends are well reproduced.  For example, the respective
$c/a$ ratios for the two codes are 1.085 and 1.083 at the constrained
$a$, and 1.047 and 1.044 at the equilibrium $a_0$.)
For each value of the applied field, we computed the relaxed structure
subject to the epitaxial constraints, and computed the strain
$\eta_3=(c-c_0)/c_0$, where $c$ is the $z$ lattice
constant at the given field and $c_0$ is the zero-field value.

The main panel of Fig.~\ref{fig:comp-comp} shows the results in
the range up to 500\,MV/m, together with a
comparison with the experimental data of Grigoriev
{\it et al.}\cite{grigoriev-prl08} on PZT 20/80, while the inset
shows the computed behavior over the full range.
The higher fields, above $\sim$500\,MV/m, are still outside the range
that is achievable experimentally, even with pulsed-field techniques.
We find that the strain increases monotonically over the entire range
up to 1400\,MV/m, with only a small tendency to begin saturating at
the highest fields.  In the range up to 500\,MV/m, the results look
very nearly linear.  We obtain an excellent fit to the results with
a simple quadratic form $\eta(E) = a_{1}E + a_{2}E^{2}$ (where
$\eta_3=\eta$ and $E_3=E$ henceforth).
We obtain fitted values of $a_1\simeq0.063-0.071$\,m/GV and
$a_2\simeq-0.01$\,(m/GV)$^2$, where the quoted range
reflects the choice of computer code.

\begin{figure}
\includegraphics[width=3.4in]{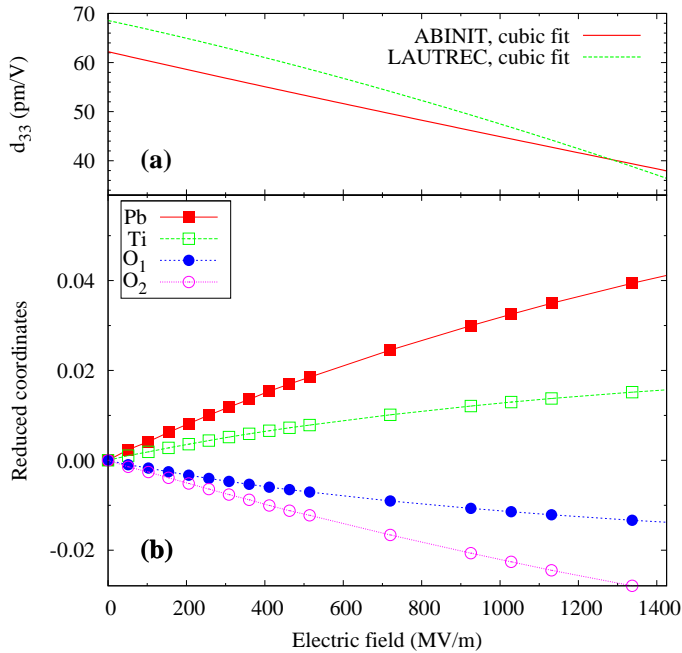}
\caption{\label{fig:epid33}
  Variation of properties of tetragonal PbTiO$_3$ with electric field
  under in-plane epitaxial constraint.
  (a) Piezoelectric coefficient $d_{33}$ (calculated with both codes).
  (b) Out-of-plane displacements of ions from zero-field positions
  (calculated with ABINIT).}
\end{figure}

The slopes of the curves in Fig.~\ref{fig:comp-comp} are, of course,
related to the piezoelectric response of the material.
At finite field it is possible to define two slightly different
piezoelectric coefficients $\tilde{d}_{33}=d\eta/dE$ and
$d_{33}=dc/dV$, where $V=cE$ is the potential drop across a unit
cell; from $d\eta=c^{-1}dc$ it follows that the definitions are related by
$d_{33}^{-1}=\tilde{d}_{33}^{-1}+E$.
We report $d_{33}$ values here, consistent with the conventions
of Ref.~\onlinecite{grigoriev-prl08}. However, the two definitions coincide
at $E=0$ and differ by only about 2\% at 500\,MV/m, so the
difference is not significant in what follows.
The results are presented in Fig.~\ref{fig:epid33}(a);
we find a monotonically decreasing trend with electric field for
both ABINIT and LAUTREC. However, the fall-off is quite slow, decreasing
by only about 2-3\% up to a field of 50\,MV/m.

\begin{figure}
\includegraphics[width=3.4in]{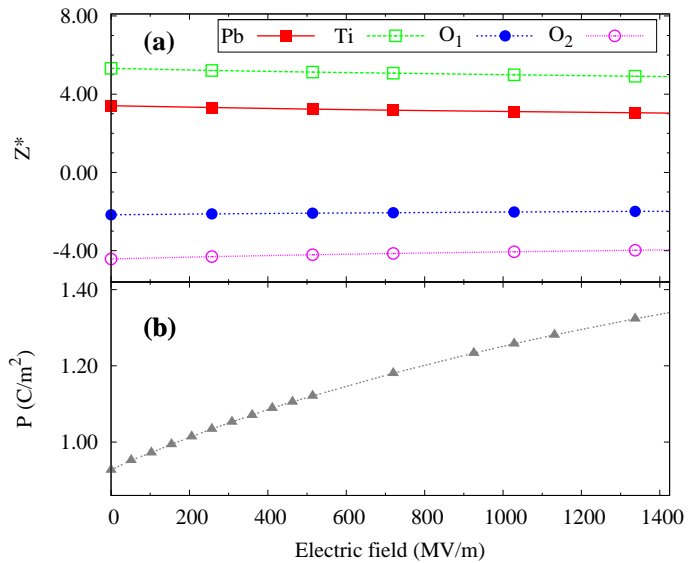}
\caption{\label{fig:polz}
  Variation of properties of tetragonal PbTiO$_3$ with electric field
  under in-plane epitaxial constraint, calculated with ABINIT.
  (a) Born effective charges.
  (b) Polarization.}
\end{figure}

Other properties of PT under applied field show a similar,
nearly linear behavior.  In Fig.~\ref{fig:epid33}(b) we present the
variations of the internal coordinates as a function of applied field
(here computed using ABINIT, but similar results are obtained with
LAUTREC).
As expected, Pb and Ti ions, being positively charged, displace along
the direction of the applied electric field, while O atoms displace
in the opposite direction, thus increasing the ferroelectric mode
amplitude.  The Born effective charges $Z^*$, shown in
Fig.~\ref{fig:polz}(a), are almost independent of field, with only
a mild reduction in their magnitudes with
increasing field. However, at zero field the $Z^*$ values under
the epitaxial constraint
are noticeably smaller in magnitude than their free-stress
counterparts.\cite{zhong-prl94}
Since the $Z^*$'s are almost constant, we expect the electric
polarization to increase with field in proportion to the
displacements shown in Fig.~\ref{fig:epid33}(b), and
Fig.~\ref{fig:polz}(b) shows that this is indeed the case.
A satisfactory fit of the form $P(E) = c_{0} +
c_{1}E + c_{2}E^{2}$ is obtained with $c_0 = 0.94$ C/m$^2$, $c_1 =
0.39$ C/(GV)m and $c_2 = -0.072$ C/(GV)$^2$.
None of these quantities show any anomalous change with
electric field.

We now return to a discussion of the strain response.
Our theory is in qualitative agreement with the recent
pulse-field data,\cite{grigoriev-prl08} which show a nearly linear
behavior similar to that predicted theoretically.  The theory has a
somewhat larger slope; our zero-field $d_{33}$ of $\sim$68\,pm/V
can be compared with their 45\,pm/V.
The agreement seems reasonable given that the materials are
different (PT vs.~PZT 20/80).
Hints of a hump-like nonlinearity in the experimental data of
Ref.~\onlinecite{grigoriev-prl08} around 200\,MV/m
are not supported by the theory; if
such a feature is really present, we would argue that it must arise
from extrinsic effects not considered by the theory.

Our zero-field $d_{33}$ is also significantly larger than the value
of 45\,pm/V
obtained by the Landau-Devonshire theory for PZT 20/80,\cite{chen-jap03}
which is in reasonable agreement with the zero-field value
of Grigoriev {\it et al}.
(For detailed comparisons, it should be kept in mind that the
misfit of $-$0.5\% reported in Ref.~\onlinecite{chen-jap03}
is only about half of that in Ref.~\onlinecite{grigoriev-prl08}.)
However, our
theoretical curves and the experimental data of
Grigoriev {\it et al.}\ both show a much slower saturation of the
strain response with field than was found by the
Landau-Devonshire approach, where $d_{33}$ was predicted to fall
by about 15\% already at 50\,MV/m for PZT 20/80 with epitaxy
constraint.  Our $d_{33}$ falls by only about 2-3\% over the
same range.
The lack of saturation means that we predict large strains of about
3.0\% at fields of around 500\,MV/m, and we also predict a large
polarization of about 1.10\,C/m$^2$ 
at such fields.  Possible reasons
for the discrepancies between our results and those of the
Landau-Devonshire theory will be discussed further in
Sec.~\ref{sec:discuss}.

The Landau-Devonshire theory\cite{chen-jap03} also provided a
zero-field value of 31\,pm/V, falling by about 10\% at 50\,MV/m,
for pure PT on an ST substrate.
Recall that our calculations above were carried out at an
in-plane lattice constant that was reduced by 1.14\% relative
to that of PT in order to model the strain state of PZT 20/80
in coherent epitaxy on a ST substrate.  Therefore, in
order to obtain a more direct comparison for the pure PT case,
where the epitaxy with the ST substrate is almost perfect,
we repeated our calculations with the in-plane lattice constant
fixed to the theoretical equilibrium lattice constant of PT,
finding a zero-field $d_{33}$ of 50\,pm/V.  This is reduced
somewhat from our value of 68\,pm/ at $-$1.14\% misfit, but still
quite a bit larger than their value of 31\,pm/V.  Again, we find a
very slow saturation (even slower that for the smaller in-plane
lattice constant), in disagreement with the Landau-Devonshire
theory.

\subsection{\label{sec:fur-cal}Application of negative pressure and removal of
epitaxy constraint}

In this section we present the results of further investigations
of the behavior of the system under modified elastic boundary
conditions that are not directly relevant to the thin-film experiments.
These investigations are motivated in part by previous {\it ab initio}
calculations of Tinte {\it et al.},\cite{tinte-prb03} who showed that
the $c/a$ ratio and polarization of PT undergo an anomalous and
strongly nonlinear variation as a function of an artificial applied
negative pressure.  The $c/a$ ratio was found to increase gradually
until the negative pressure reached about $-$4.8\,GPa, where the
$c/a$ ratio rapidly increases to $\sim$1.20, with a subsequent slower
increase up to $\sim$1.25 at $-$7\,GPa.  Neither an electric field
nor an epitaxy constraint were applied in that calculation, but the
results nevertheless hint at a possible supertetragonal state of
PT that might be accessed under unusual boundary conditions.
We also note that negative pressure is sometimes used to simulate the
effect of chemical substitution, and that such an approach may be
relevant here for providing hints about the behavior of PZT solid
solutions.  Zr and Ti belong to the same column of the Periodic Table,
but Zr has a larger effective radius and will therefore tend to expand
the lattice constant of PZT relative to pure PT; from the point
of view of a Ti ion, this could produce a similar effect as is obtained
from application of a negative pressure to pure PT.

Here, we explore the strain response as a function of both electric
field and negative pressure, with and without the epitaxial constraint.
The exercise is carried out in three steps.
First, we apply a negative pressure in the presence of the
epitaxial constraint, obtaining
the strain response as a function of electric field.
(In this case only the $zz$ component of the
stress tensor is relevant, and we could equally well say that we
are applying a tensile uniaxial stress along $c$.)
Second, we remove
the negative pressure and repeat the calculations presented in
Sec.~\ref{sec:cal-expt}, but without epitaxial constraint,
so that the crystal is free to relax its in-plane lattice constants
in response to the applied field.  Finally, we combine
the two steps described above and study the behavior as a function
of electric field and negative hydrostatic pressure in the absence of
any epitaxial constraint.  Some results of this kind, both at zero and
negative pressure, have already been presented in
Ref.~\onlinecite{stengel-np09}.

\begin{figure}
\includegraphics[width=3.4in]{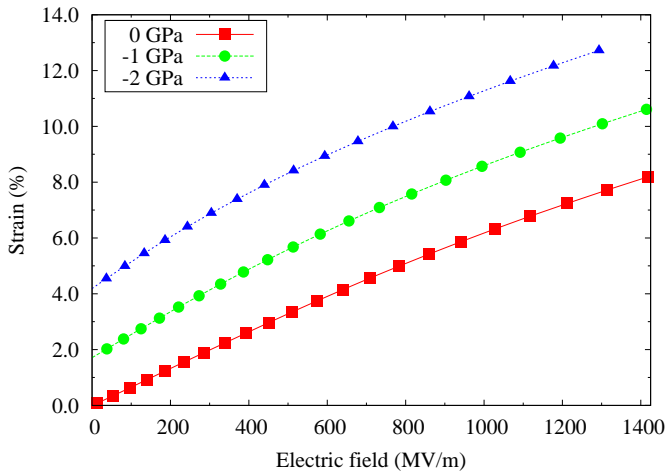}
\caption{\label{fig:con-pr}
  Field-induced strain, relative to zero-field zero-pressure state,
  for tetragonal PbTiO$_3$ under in-plane epitaxial constraint,
  calculated with LAUTREC.}
\end{figure}

Our present results for the application of negative pressure
in the presence
of the epitaxial constraint are shown in Fig.~\ref{fig:con-pr}, where
the strain plotted on the vertical axis is defined as relative to the
$c$ lattice constant at zero pressure and zero field. In this
and subsequent figures we present calculations performed
with the LAUTREC package.
The zero-pressure curve duplicates the data presented in
Fig.~\ref{fig:comp-comp}, while the ones at $-$1 and $-$2
show the enhancement in the strain caused by the negative pressure,
which shifts the curves by about 2\% per GPa.
Otherwise they look rather similar, except that there is slightly
more nonlinearity in the curves as the pressure becomes more
negative.

\begin{figure}
\includegraphics[width=3.4in]{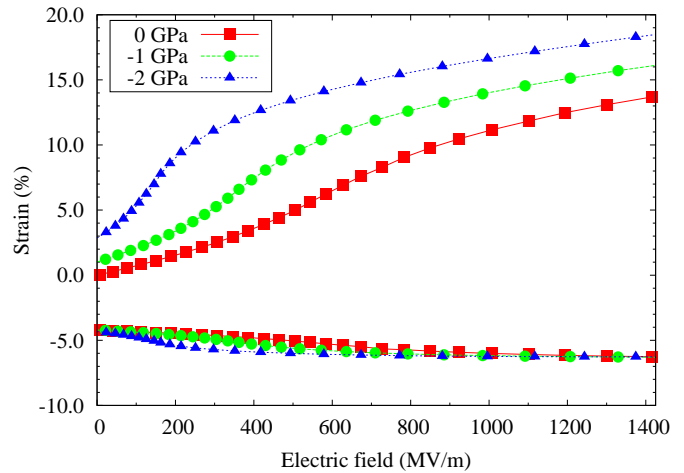}
\caption{\label{fig:nocon-pr}
  Field-induced strain, relative to out-of-plane lattice constant
  of zero-field zero-pressure state,
  for tetragonal PbTiO$_3$ under free-stress (0 GPa) and negative-pressure
  ($-$1 and $-$2\,GPa) boundary conditions, calculated with LAUTREC.
  Top curves: out-of-plane strain.
  Bottom curves: in-plane strain.}
\end{figure}

\begin{figure}
\includegraphics[width=3.4in]{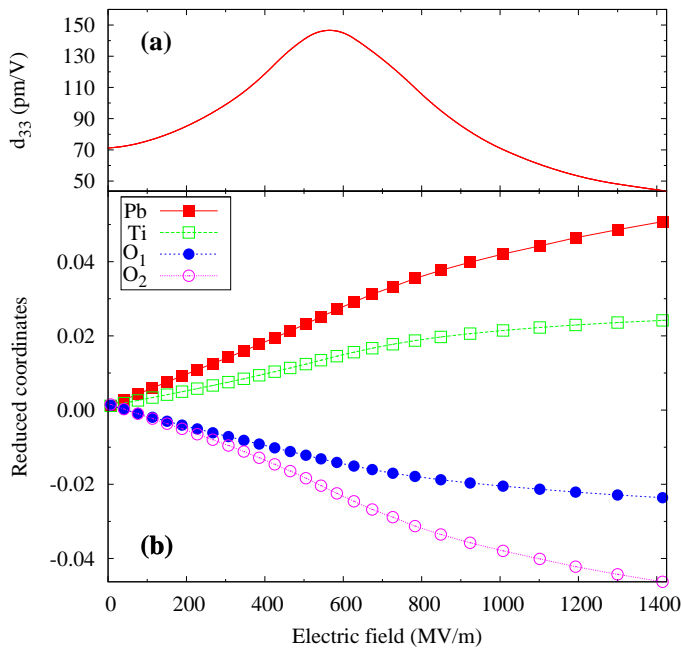}
\caption{\label{fig:ionpos_rel}
  Variation of properties of tetragonal PbTiO$_3$ with electric field
  under free-stress boundary conditions, calculated with LAUTREC.
  (a) Piezoelectric coefficient $d_{33}$.
  (b) Out-of-plane displacements of ions from zero-field positions.}
\end{figure}

The removal of the epitaxial constraint causes a much more
drastic change in behavior, however, as shown by the square
symbols (red curves) in Fig.~\ref{fig:nocon-pr}.  Both in-plane
and out-of-plane strains are defined relative to the $c$ lattice
constant at zero field.  As expected, the enhanced polarization and
enhanced tetragonality induced by the field cause the out-of-plane
lattice constant to grow, and the in-plane one to shrink, with
increasing field.  However, this variation shows an anomalous
behavior: the change in strain starts out slowly at lower fields,
accelerates and occurs most rapidly at a characteristic field
$E_{\rm anom}\simeq550$\,MV/m, and then slows again at higher fields.
This anomalous behavior is most evident in the curve for the
out-of-plane strain, but is also visible for the in-plane strain.
The anomaly also shows up clearly in the behaviors of the
piezoelectric coefficient and internal displacements, presented in
Fig.~\ref{fig:ionpos_rel}(a) and (b) respectively.

Finally, the remaining curves in Fig.~\ref{fig:nocon-pr} show
how this behavior evolves as a negative pressure is applied.
The anomaly becomes more pronounced as the pressure becomes more
negative, with $E_{\rm anom}$ shifting
to lower fields of about 375 and 150\,MV/m at $-$1 and $-$2\,GPa
respectively.
(This behavior was also presented in Ref.~\onlinecite{stengel-np09}
in terms of reduced electrical coordinates.)
Very large strains (with $c/a$ ratios approaching
1.25) occur at the largest fields and strongest negative pressures
considered.  The behaviors are similar to those already reported in
Ref.~\onlinecite{tinte-prb03} as a function of negative pressure
alone, but here we find that the application of an electric
field works cooperatively with the negative pressure to induce
the crossover into the supertetragonal state.\cite{stengel-np09}

Returning to the zero-pressure results, we can now compare these
more directly with previous works in which the epitaxy constraint
was absent.  As was shown in Fig.~\ref{fig:ionpos_rel}(a), we find
that $d_{33}$ starts at about 70\,pm/V at zero field, increases up
to about 140\,pm/V near $E_{\rm anom}=550$\,MV/m, and then falls
again at higher fields.  The results of Ref.~\onlinecite{sai-prb02}
instead show $d_{33}$ starting at 39\,pm/V and decreasing
monotonically as the field is increased.
The discrepancy with respect to our results may be attributed
to the more approximate methods used in Ref.~\onlinecite{sai-prb02},
where the electric field was applied in an approximation in which
extra forces appear on the atoms in proportion to the Born effective
charges computed at zero field.\cite{sai-prb02,fu-prl03}
The Landau-Devonshire
results of Ref.~\onlinecite{chen-jap03} also show $d_{33}$ decreasing
monotonically with field, but starting from a zero-field value of
79\,pm/V that is closer to ours.  The comparison between their work and
ours deserve further comment, as will be provided in the next section.

\subsection{Discussion}
\label{sec:discuss}

In this section we discuss some of the factors that need to be
considered when comparing our theoretical calculations with previous
theory and experiment.  We focus first on three important
considerations, namely, the elastic boundary conditions (epitaxial
vs.\ free-stress), the reference in-plane lattice constant, and the Zr
content.  We then discuss several other factors that may also play a
significant role. Some comparisons have already been made along these
lines in Sec.~\ref{sec:cal-expt}, but we concentrate here on possible
mechanisms and physical explanations for the observed trends.

\subsubsection{Reference in-plane lattice constant}
\label{sec:lat-con}

When applying the epitaxial constraint, we have a choice of which
in-plane lattice constant $a$ to use for the comparison.  As explained
earlier, most of our results are reported for a value of $a$ that is
1.14\% smaller than the equilibrium lattice constant of PT,
to facilitate comparison with
the experiment of Grigoriev {\it et al.}\cite{grigoriev-prl08} on PZT 20/80.
If instead we repeat the calculations by fixing $a$ to be that of
tetragonal PT at its theoretical equilibrium lattice constant,
as mentioned earlier near the end of Sec.~\ref{sec:cal-expt},
we find a zero-field $d_{33}$ of about 50\,pm/V,
to be compared with the value of about 68\,pm/V obtained at the
smaller lattice constant with LAUTREC code.
This trend is understandable since one may expect a larger $d_{33}$
in a more tetragonal material.  It is also consistent with the
trend shown in Fig.~4(a) of Chen {\it et al.}, where $d_{33}$ increases
from 39 to 58\,pm/V when the epitaxy constraint is made
1\% more compressive.

\subsubsection{Epitaxial vs.~free-stress boundary conditions}

Our work reinforces the expectation that the choice of epitaxial
vs.\ free-stress elastic boundary conditions plays a crucial role
in the piezoelectric response.  This is true already at zero field,
but the effect becomes enormous for elevated fields.

At zero field, one expects the free-stress $d_{33}$ to exceed the
epitaxial one, other things being equal.  This can be understood
as follows.  If we start from the relaxed tetragonal
ferroelectric PT crystal and apply an enhancing electric
field along $z$ with the in-plane $a$ fixed, we can decompose the
response according to a two-step process.  First, we apply the field while
allowing $a$ and $c$ to relax.  We expect the degree of tetragonality to
increase along with $P$, so that $a$ shrinks while $c$ grows, as
confirmed by Fig.~\ref{fig:nocon-pr}.  Second, we enlarge $a$ back
to its zero-field value, again while allowing $c$ to relax; since
the Poisson ratio is positive, this should cause $c$ to shrink.
Thus, $d_{33}$ should be smaller in the epi-constrained case.
This is true in the work of Chen {\it et al.}\cite{chen-jap03},
where for example the zero-field $d_{33}$ decreases from
79 to 31\,pm/V for pure PT, and from 87 to 45\,pm/V
for  PZT 20/80, when going from free-stress to epi-constrained
boundary conditions.  It is also confirmed in our calculations
on pure PT; we find that the zero-field $d_{33}$
decreases from 71 to 50\,pm/V (LAUTREC values) in going from the
free-stress to the epi-constrained case.

As indicated earlier, we find a drastic difference in the \emph{non-linear}
response when we remove the epitaxy constraint, with the smooth
decrease of $d_{33}$ in Fig.~\ref{fig:epid33}(a) replaced by the rapid
increase and peak around 550\,MV/m in Fig.~\ref{fig:ionpos_rel}(a).
This peak is clearly a signature of the crossover into the
supertetragonal state.

\subsubsection{Zr content}

Of course, it is important to keep track of the differences
between pure PT and PZT with different Zr concentrations, as we
have tried to do consistently above.  As is well known, PZT is
often preferred as a more practical material for experimental
purposes and for applications because of reduced leakage currents
and other beneficial properties.  The work of Chen
{\it et al.}\cite{chen-jap03} reports $d_{33}$ increasing from
79 to 87\,pm/V in going from pure PT to PZT 20/80 in the
free-stress case, and from 31 to 39\,pm/V in the epi-constrained
case at zero misfit.
If this trend is correct, it is in the wrong direction to explain
the difference between our theory and the experiment of
Grigoriev {\it et al.}\cite{grigoriev-prl08}, since our $d_{33}$
values for pure PT are larger than theirs for PZT 20/80.
Other factors may be responsible for this discrepancy, as discussed
in the next subsection.  Ideally it would be advantageous
if the pulsed-field experiments could be carried out on a series of
samples of varying Zr content so that an extrapolation could be
made to the pure-PZ case, allowing a more direct comparison with
theory.

\subsubsection{Other factors}

Here we comment briefly on a number of other factors that might play a
role when comparing our results with experiment.  Our theory is purely
a zero-temperature theory, and also completely neglects the effects of
disorder arising from Zr configurations in PZT, or from defects such
as oxygen vacancies even in pure PT.  Both disorder and thermal
fluctuations may tend to cause the local polarization direction to
fluctuate about the global tetragonal axis, and it is well known that
polarization rotation can generate large electromechanical responses
in this class of materials.\cite{fu-np00} At larger length scales, it
is also possible that the samples might not be in a perfect
single-domain state, with the incomplete switching being caused by
defects, roughness, or other imperfections in the thin-film samples.
In such a case, the intrinsic piezoelectric response would be
underestimated, while extrinsic contributions associated with
domain-wall motion might also be present.  For ultrathin films,
depolarization effects and the influence of the perovskite-electrode
interfaces \cite{nmat_2009,capa_2009} will also play an important
role.  We expect that these issues will be clarified as improved
methods of sample preparation become available, and as systematic
studies are done to see how the piezoelectric responses depend on film
thickness and other properties.

$\phantom{}$

\section{Summary}
\label{sec:summary}

In summary, we have studied the piezoelectric response
of PbTiO$_3$ to a polarization-enhancing electric field applied
along the tetragonal axis under several kinds of
mechanical boundary conditions.
In the epitaxially constrained
regime we find hardly any saturation of the piezoelectric coefficient
$d_{33}$ up to a field of 500\,MV/m,
in agreement with recent experimental measurements and in contrast
with the predictions of Landau-theory expansions.
With the removal of the epitaxial constraint
we find a remarkable non-linear effect, with $d_{33}$ rising to twice
its zero-field value at $\sim$550\,MV/m and then decreasing again for higher
applied fields.

The comparison between ABINIT and LAUTREC shows fairly good agreement.
Some quantitative discrepancies do exist (e.g., in equilibrium
lattice constants); these can be traced to the sensitivity of
PbTiO$_3$ to the choice of pseudopotentials.  Nevertheless, the
trends in the two calculations are very similar.
Moreover, extensive tests show that the dissimilarities in the
finite-field techniques (constrained-$E$ with Berry phase in ABINIT
vs.\ constrained-$D$ with Wannier functions in LAUTREC) have little or
no impact
on the calculated properties, once the subtleties in the treatment of the
electrical and strain variables have been properly accounted for.

Quantitative comparisons with experiment still present a formidable
challenge, in part because it remains difficult to characterize the
precise sample conditions underlying a given electrical
measurement.  However, significant progress has now been made on
the theoretical side, and experimental control of film properties
continues to improve.  Thus, we hope that direct comparisons
between theory and experiment regarding nonlinear piezoelectric
behavior will provide an increasingly fruitful avenue for future
investigations.


\acknowledgments

We thank A.~Grigoriev and K.M.~Rabe for useful discussions.
The work was supported by ONR Grant N00014-05-1-0054.

\bibliography{paper}

\begin{thebibliography}{32}
\bibitem[{LAU()}]{LAUTREC}
\bibinfo{note}{LAUTREC is a massively parallel ``in-house'' electronic
  structure code.}
\bibitem[{exp()}]{explan-pulay}
\bibinfo{note}{The Pulay error in the stress tensor was compensated by applying
  a fixed hydrostatic pressure of $-$0.4 GPa during the structural relaxations
  performed with LAUTREC. No correction was used in ABINIT, where the error was
  found to be smaller than our accuracy threshold.}
\end{thebibliography}

\end{document}